\documentclass[twocolumn]{aastex7}

\begin{document}

\title{Stars Born in the Wind: M82’s Outflow and Halo Star Formation}

\author[0000-0002-8406-0136]{Vaishnav V. Rao}
\affiliation{Department of Astronomy, University of Michigan, 1085 S. University Ave, Ann Arbor, MI 48109-1107, USA}
\email[show]{vvrao@umich.edu}

\author[0000-0003-2599-7524]{Adam Smercina}\thanks{NHFP Hubble Fellow}
\affiliation{Space Telescope Science Institute, 3700 San Martin Dr., Baltimore, MD 21218, USA}
\email{asmercina@stsci.edu}

\author[0000-0002-5564-9873]{Eric F.\ Bell}
\affiliation{Department of Astronomy, University of Michigan, 1085 S. University Ave, Ann Arbor, MI 48109-1107, USA}
\email{ericbell@umich.edu}

\author[0000-0001-7531-9815]{Benjamin Williams}
\affiliation{Astronomy Department, University of Washington, Box 351580, U.W. Seattle, WA 98195-1580, USA}
\email{benw1@uw.edu}

\author[0000-0002-1264-2006]{Julianne J.\ Dalcanton}
\affiliation{Astronomy Department, University of Washington, Box 351580, Seattle, WA 98195-1580, USA}
\affiliation{Center for Computational Astrophysics, Flatiron Institute, 162 Fifth Avenue, New York, NY 10010, USA}
\email{jdalcanton@flatironinstitute.org}

\author[0000-0002-8084-8612]{Andrew Dolphin}
\affiliation{Raytheon, Tucson, AZ 85756, USA}
\affiliation{Steward Observatory, University of Arizona, Tucson, AZ 85721, USA}
\email{adolphin@rtx.com}

\author[0000-0002-2545-1700]{Adam Leroy}
\affiliation{Department of Astronomy, The Ohio State University, 140 West 18th Avenue, Columbus, OH 43210, USA}
\email{leroy.42@osu.edu}

\author[0000-0003-2325-9616]{Antonela Monachesi}
\affiliation{Departamento de Astronom\'{i}a, Universidad de La Serena, Avda. R\'{a}ul Bitr\'{a}n 1305, La Serena, Chile}
\email{amonachesi@userena.cl}

\author[0000-0001-6380-010X]{Jeremy Bailin}
\affiliation{Department of Physics and Astronomy, University of Alabama, Box 870324, Tuscaloosa, AL 35487-0324, USA}
\email{jbailin@ua.edu}

\author[0000-0001-6982-4081]{Roelof S. de Jong}
\affiliation{Leibniz-Institut f\"{u}r Astrophysik Potsdam (AIP), An der Sternwarte 16, 14482 Potsdam, Germany}
\email{rdejong@aip.de}

\author[0000-0003-4793-7880]{Fabian Walter}
\affiliation{Max-Planck Institut für Astronomie, Königstuhl 17, D-69117, Heidelberg, Germany}
\email{walter@mpia.de}

\begin{abstract}

Starburst galaxies, like M82, launch kiloparsec-scale galactic outflows that interact with the circumgalactic medium (CGM) in complex ways. Apart from enriching the CGM with metals and energy, these outflows may trigger star formation in the halo- either by driving shocks into the CGM or transporting cold, star-forming gas. To investigate such processes, we analyze the star formation history (SFH) of the Southern Arcs—arc-like stellar features located $\sim5$\,kpc from M82’s star-forming disk along the minor axis—using \textit{Hubble Space Telescope} Wide Field Camera 3 photometry. From resolved stellar populations, we derive SFHs over the last $\sim500$\,Myr, finding that $\sim85\%$ of the stellar mass formed between $\sim150$ and $\sim70$\,Myr ago, followed by a brief pause, with the remaining $\sim15$\% forming since $\sim30$\,Myr ago. The two stellar populations are co-spatial on scales of at least $\sim200\,$pc. The timing of the $\sim100$\,Myr burst aligns with star formation in the M82 disk and the age distribution of its star clusters, suggesting a causal link between the disk starburst and halo star formation. We explore two mechanisms that could explain these observations. In the first, shocks driven by the interaction between hot outflowing gas and cooler CGM material compress dense clouds, triggering collapse and star formation. In the second, stars form directly within massive, cool clouds associated with the outflow. As these clouds move ballistically through the halo, subsequent interactions with tidal debris may trigger additional star formation, producing the observed episodic structure.

\end{abstract}



\section{Introduction} \label{sec:intro}
Galactic outflows, commonly observed across the universe, play a crucial role in enriching the circumgalactic (CGM) and intergalactic medium (IGM) with metals and energy \citep{Veilleux2005, Borthakur2013}. These outflows are typically driven by hot gas bubbles formed by numerous supernovae (SNe) within a galaxy, which expand outward and entrain dust along with cold and warm gasses \citep[e.g.,][]{C&C1985}. Although launched on sub-parsec scales, they can have dramatic impacts on the large-scale galactic environment, as seen in high-resolution simulations \citep[e.g.,][]{Nelson2019}. In these models, stars can form within the outflow itself, where regions are compressed to densities above the star-formation threshold \citep[e.g.,][]{Yu2020}. Observationally, indirect spectroscopic evidence—such as complex emission line decomposition—supports the presence of multiple star-forming phases within these outflows \citep[e.g.,][]{Maiolino2017, Gallagher2019}.

Another possible mode of star formation occurs when shock fronts produced at the interface of the outflow and the surrounding medium encounter a molecular cloud, causing it to collapse and form stars. This kind of triggered star formation has been observed, for example, in the ``inner" and ``outer" filaments of Centaurus A \citep[Cen A;][]{Rejkuba2002, GK&W2010, Crockett2012}. Despite the fact that this is due to an AGN jet, the physical mechanisms are similar to star formation-driven winds. Scenarios of shock fronts inducing star formation in molecular clouds have been reproduced in simulations \citep[e.g.,][]{Gardner2017, Kinoshita2021a} and there is growing observational evidence for star formation triggered by supernovae, ionization fronts, cloud-cloud collisions, and other shocks in the interstellar medium of our own galaxy \citep[e.g.,][]{H&D2005, Furukawa2009, Kinoshita2021b}.

Despite these insights, key questions remain unresolved. Where and how do starburst-driven outflows trigger star formation? Are these outflows intermittent, with measurable duty cycles \citep[e.g.,][]{Lee2009}? How are they connected to the starburst itself? Addressing these questions remains difficult as most known starburst-driven outflows occur in galaxies that are too distant for detailed study of their stellar populations and physical environments \citep[e.g.,][]{Leroy2015}.

\begin{figure*}
    \centering
    \includegraphics[width=\linewidth]{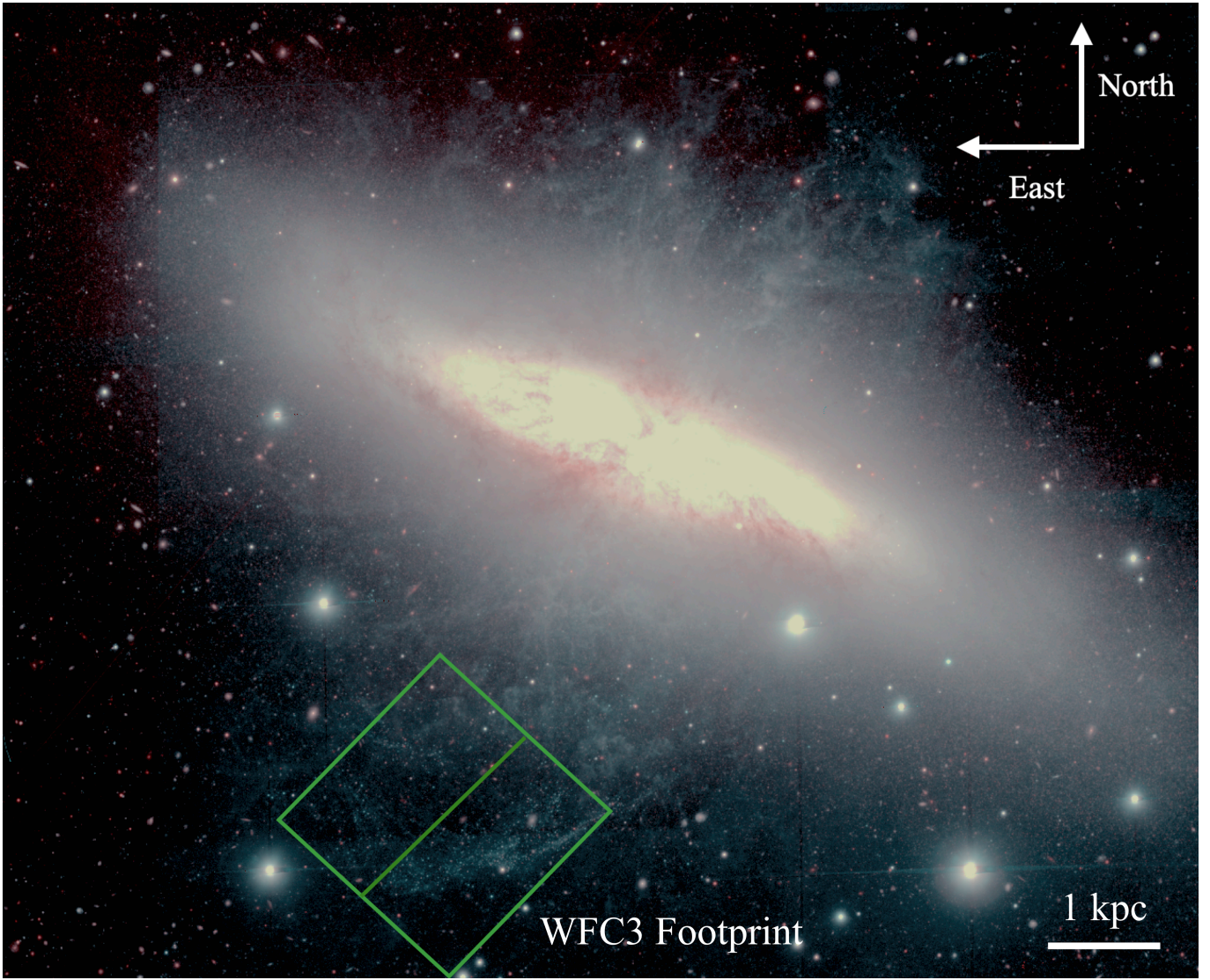}
    \caption{\textit{Hubble Space Telescope} (HST) Wide Field Camera 3 (WFC3) footprint of this work overlaid on the $gri$ color image of M82 from Subaru Hyper Suprime Cam (HSC) \citep{Smercina2020}.}
    \label{fig:wfc3-footprint}
\end{figure*}

M82 stands out as the best galaxy in the local universe (at a distance of 3.6\,Mpc; \citealt{Dalcanton2009}) for detailed studies of starburst-driven outflows. This edge-on galaxy hosts a prototypical nuclear starburst, which drives strong, multiphase, bipolar outflows (``superwinds") along its minor axis \citep{L&S1963, O&M1978, Bland&Tully1988}. The starburst activity is widely attributed to a close encounter with Milky Way-mass M81 within the last $\sim1$\,Gyr \citep[e.g.,][]{deGrijs2001, RM2011, Mayya2006, Li2015}. Ongoing interactions with both M81 and the LMC-like NGC 3077 further complicate its environment, generating significant tidal debris composed of streams and filaments of HI gas that extend out to approximately 10\,kpc along the minor axis \citep{Yun1994, Okamoto2015, deBlok2018, Smercina2020}. This debris forms part of an extensive HI structure that bridges M82 with M81 and NGC 3077 \citep[see Fig. 2 of][]{deBlok2018}, making it an exceptional target for studying the full impact of outflows on galactic and intergalactic environments.

The multiphase outflows in M82 \citep{Leroy2015} have been studied across the electromagnetic spectrum, with observations spanning cold atomic HI and molecular gas \citep{Walter2002, Salak2013, Beirao2015, Martini2018, Fisher2025}, warm-ionized gas traced by H$\alpha$ \citep{McKeith1995, Westmoquette2009}, hot gas in X-rays \citep{Watson1984, Bregman1995, Strickland1997, Lopez2020}, and entrained dust observed in UV, IR, and sub-millimeter wavelengths \citep{Hoopes2005, L&R2009, Kaneda2010, Roussel2010}. These studies collectively reveal a complex picture: the hot gas outflow, tightly confined by CO and HI, extends far into the halo, merging with the surrounding tidal debris. Meanwhile, the cold component of the outflow does not reach beyond $\sim4$\,kpc from the disk, as it stalls and ultimately falls back as a “cold fountain” \citep{Leroy2015}. 

It is then clear that studying stellar populations that are spatially coincident with the outflow could provide a unique opportunity to investigate how starburst-driven winds affect halo star formation. In this context, arcs of stars south of M82 -- henceforth the Southern Arcs --- are of particular interest. To our knowledge, these features were first highlighted by \citet{Sun2005} and were subsequently studied by \citet{Davidge2008}, \citet{Okamoto2019}, and \citet{S&S2024}. The arcs lie at a distance of $\sim3.6$\,Mpc  --- similar to the distances to M82 and M81 \citep{Dalcanton2009, S&S2024}, and are situated where the hot gas outflow is expected to interact with the tidal debris in the circumgalactic medium (CGM). Ground-based observations by \citet{Davidge2008} suggested that the youngest stars in the arcs have ages $\sim50$\,Myr, comparable to other features in the M81 Group such as Holmberg IX, the Garland, and the Arp Loop \citep[e.g.,][]{S&M2001, Makarova2002}. More recent work by \citet{S&S2024} proposed slightly older ages for the arc populations, fitting them to 100–160\,Myr isochrones. Both studies suggest that the southern outflow’s interaction with the surrounding HI tidal debris could have induced the arcs' star formation. However, no detailed SFH analysis of this region has been conducted to date. We have therefore targeted the sparse halo region $\sim5$\,kpc from the disk center, containing this young population using the \textit{Hubble Space Telescope}'s (HST) Wide Field Camera 3 (WFC3) (Fig. \ref{fig:wfc3-footprint}) to obtain deep photometry of its stars. By deriving detailed SFHs, we aim to investigate the time evolution of M82’s outflow and its role in linking star formation across galactic scales. 

In Section \ref{sec:data}, we describe the photometry of the Southern Arcs and characterize its completeness and biases. We also introduce the archival ACS Nearby Galaxy Survey Treasury \citep[ANGST;][]{Dalcanton2009} dataset, which covers the broader M82 field. In Section \ref{sec:panoramic-view}, we combine the Southern Arcs and ANGST datasets to generate a panoramic view of M82's young and intermediate-age stellar populations. Section \ref{sec:sfh} details the methods used to measure the star formation histories (SFHs) of the Southern Arcs and presents the results. In Section \ref{sec:spat-dist}, we examine the spatial distribution of different stellar populations in the Southern Arcs field and place constraints on their dynamics. In Section \ref{sec:discussion}, we place these results in the context of stellar populations in M82’s disk, multi-wavelength maps of its outflow, and the connection between the Southern Arcs and the starburst-driven wind. Here, we also propose two plausible mechanisms for the formation of the Southern Arcs. Finally, we summarize our findings in Section \ref{sec:summary}.

\section{Data}\label{sec:data}
\subsection{Southern Arcs: Observations}

The Southern Arcs field was observed with HST’s UVIS channel of the Wide Field Camera 3 (WFC3) between January 2021 and February 2022. Observations were taken in the F475W and F814W filters as part of HST Cycle 28 program GO-16185 (PI: Adam Smercina). The field was observed over four single-orbit visits, with three exposures per orbit. The first visit was split, with a two-point dither in F814W and a single exposure in F475W. Subsequent visits each used a single filter, with a three-point linear dither (two visits in F814W and one in F475W). The dither spacing was set to 2.98\arcsec\ to achieve sub-pixel sampling of the point-spread function and to cover the WFC3 UVIS chip gap. The position angle (PA) of the observations was 49.8\degr. The field of view is 162\arcsec\ $\times$ 162\arcsec, centered on (RA, Dec) = (09:56:21.030, +69:36:50.90), with a plate scale of 0.04 arcsec per pixel. The total exposure time and 50\% completeness depth (measured from artificial star tests; Section \ref{sec:ast}) in each filter are summarized in Table~\ref{tab:obs}. These HST data can be found in MAST: \dataset[10.17909/ss76-p487]{http://dx.doi.org/10.17909/ss76-p487}

\begin{deluxetable}{lcc}
\tablecaption{WFC3 Observations of the Southern Arcs\label{tab:obs}}
\tablehead{
\colhead{Filter} & \colhead{Exposure Time (s)} & \colhead{50\% Completeness}
}
\startdata
F475W     & 3482 & 28.1 \\
F814W &  6964 & 26.5 \\
\enddata
\tablecomments{(RA,Dec)= (09:56:21.030,+69:36:50.90)}
\end{deluxetable}

\subsection{Southern Arcs: Photometry and ASTs}\label{sec:ast}
We performed point spread function (PSF) fitting on the HST pipeline-calibrated image (\texttt{flc} extension) using the latest version of the \texttt{DOLPHOT} software package \cite{Dolphin2000PHOT, Dolphin2016}. We conducted artificial star tests (ASTs) to evaluate the photometric quality, completeness, and bias of our observations. We generated an input list of 250,000 artificial stars, sampled from a realistic range of stellar population models employing the Bayesian Extinction and Stellar Tool (\texttt{BEAST}) package \citep{Gordon2016}. We then injected these artificial stars into the image and recovered them through \texttt{DOLPHOT} following the methodology established by the Panchromatic Hubble Andromeda Treasury (PHAT) program \citep{Williams2014}.

\begin{figure*}[ht!]
    \centering
    \includegraphics[width=\linewidth]{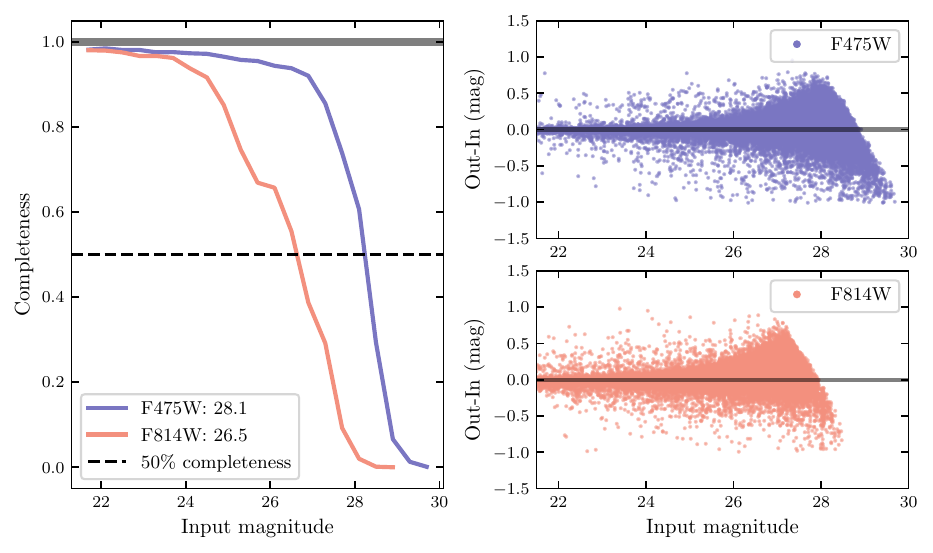}
    \caption{Completeness of photometry in the Southern Arcs as a function of input magnitude (\textbf{left}) and photometric bias or error in recovered magnitudes as a function of input magnitude for the F475W (\textbf{top right}) and F814W (\textbf{bottom right}) filters from artificial star tests (ASTs).}
    \label{fig:comp-bias}
\end{figure*}

Using the results of the ASTs, we computed completeness curves for each filter. These curves help us determine the optimal selection criteria for stellar sources. We classified sources as `good stars' (GST) if they met our criteria for \texttt{DOLPHOT}'s signal-to-noise ratio (\texttt{SNR}), crowding (\texttt{CROWD}), and sharpness (\texttt{SHARP}) parameter outputs. In the uncrowded Southern Arcs, the \texttt{CROWD} parameter was particularly useful for identifying spurious sources detected along the diffraction spikes of bright foreground stars and other photometric artifacts. We chose the smallest \texttt{CROWD} parameter that effectively removes most of these contaminating sources without significantly impacting photometric completeness. The final criteria for selecting GSTs were: \texttt{SNR} $> 4$, \texttt{SHARP}$^2 < 0.2$, and \texttt{CROWD} $< 0.25$. This resulted in 4902 GST sources. In Figure \ref{fig:comp-bias} we show the completeness curves for the GST sources along with the 50\% completeness limits for each filter. We also show the photometric bias for the GST sources in each filter.

\subsection{ANGST: Archival Dataset}
HST imaging of M82 for the ANGST survey was obtained in March 2006 by \citet{Mutchler2007} and the Hubble Heritage Team (HST Proposal 10776) using the Wide Field Channel (WFC) of the Advanced Camera for Surveys (ACS). Observations were taken in the F435W, F555W, F814W, and F658N ($H\alpha$) filters across six pointings, arranged in a 3×2 mosaic to cover M82, extending out to $\sim3.5$ \,kpc from its disk. Photometry in the F435W, F555W, and F814W filters was extracted as part of the ANGST survey and is detailed in \citet{Dalcanton2009}. We selected GSTs from this dataset using the same criteria as for the Southern Arcs: \texttt{SNR} $> 4$, \texttt{SHARP}$^2 < 0.2$, and \texttt{CROWD} $< 0.25$.

\begin{figure*}[ht!]
    \centering
    \includegraphics[width=\linewidth]{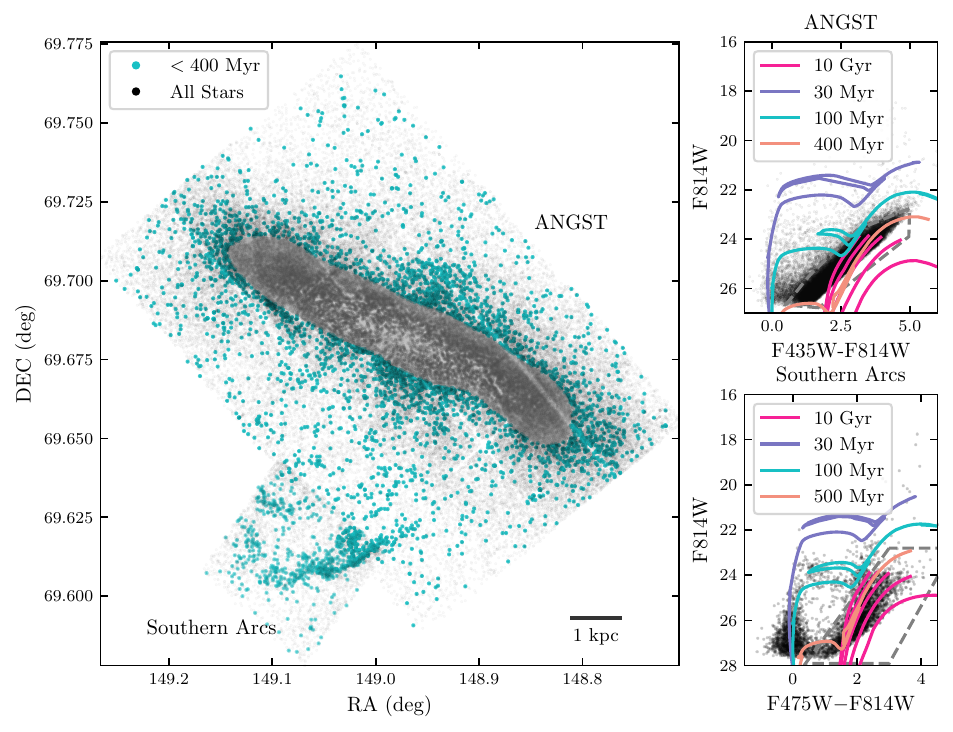}
    \caption{(\textbf{Left}) Resolved stars in the M82 field from ANGST \citep{Dalcanton2009} with the stars from the Southern Arcs field (this work). The points in cyan represent populations of stars $\lesssim 400$\,Myr, selected by excluding the red giant branch (RGB) stars from the color-magnitude diagram (CMD) of each dataset (dashed grey regions in the panels on the right) with matched depths (F814W$<26$). The M82 disk has been masked out for these stars to accentuate structures that follow the breakout of the outflow from the disk. Arc-like features in the Southern Arcs field appear to be spatial extensions of young stellar structures in ANGST that may be related to M82's starburst outflow. (\textbf{Top Right}) F814W vs. F435W-F814 CMD of `good star' (GST) sources from ANGST with some typical PARSEC \citep{Bressan2012-PARSEC} isochrones (shifted assuming $A_V=0.1$ and $(m-M)_0 =27.8$). The 10\,Gyr isochrones have metallicities, [M/H] = 0, -0.5, -1.0, -1.5, -2.0. The younger isochrones have [M/H] = 0.
    (\textbf{Bottom Right}) CMD of `GST' sources from the Southern Arcs. PARSEC isochrones  with a foreground extinction (shifted assuming $A_V=0.35$ and $(m-M)_0=27.8$) are over-plotted to broadly explain the multiple stellar populations. The 10\,Gyr isochrones ([M/H] = 0, -0.5, -1.0, -1.5, -2.0) correspond to the old, RGB halo stars. The 100\,Myr isochrone accounts for some of the fainter main sequence (MS) stars and most of horizontal branch core helium burners (HeB). The 30\,Myr isochrone accounts for the brighter MS stars, HeBs, and red supergiants. We assume an [M/H]$= -$0.25 for the $\le500$\,Myr isochrones.}
    \label{fig:m82-angst}
\end{figure*}

\section{Panoramic View of M82's Resolved Stars}\label{sec:panoramic-view}
We map the resolved stars in M82 from both the ANGST fields and the Southern Arcs field in Fig. \ref{fig:m82-angst}. While the ANGST photometry is shallow and incomplete—largely due to extreme crowding and significant dust obscuration near the disk ($A_V \sim 1$–4; \citealt{Hutton2015})—it nevertheless reveals a widespread distribution of young and intermediate-age stars ($\lesssim 400$\,Myr) throughout M82’s halo. We selected these stars from the color-magnitude diagram (CMD) by excluding regions occupied by the older red giant branch (RGB) halo stars (top right panel of Fig. \ref{fig:m82-angst}). Most notably, structures immediately to the north and south of the disk stand out, coinciding with the regions where the outflow emerges. These structures provide qualitative hints that outflow-driven star formation may be occurring near the base of M82’s multiphase winds.

At comparable F814W depth, the Southern Arcs appear strikingly rich in young and intermediate-age stars and seem to extend the stellar structures seen in ANGST. The young and intermediate-age stars in the Southern Arcs were similarly selected from the CMD (bottom right panel of Fig. \ref{fig:m82-angst}). Qualitatively, the CMD of the Southern Arcs reveals two distinct stellar populations. The first consists of RGB stars older than $\gtrsim 1$\,Gyr, likely associated with M82’s stellar halo \citep{Smercina2020, Velguth2024, Okamoto2015}, and are well represented by 10\,Gyr isochrones. The second population includes main sequence (MS) and core helium-burning (HeB) stars younger than $\lesssim 1$\,Gyr, which may have formed under the influence of M82’s outflow. This younger population is well described by a combination of a 100\,Myr isochrone—accounting for fainter MS stars and most horizontal branch HeBs—and a 30\,Myr isochrone, which traces the brighter MS stars, HeBs, and red supergiants.

\citet{Williams2011} used the ANGST dataset to derive a global star formation history (SFH) for M82 by combining SFHs from the six individual tiles. However, the results were limited by coarse time resolution and substantial uncertainties, mainly due to the severe crowding and dust extinction in M82’s disk, which constrained the precision of the inferred SFH. Given how striking the young populations are in the Southern Arcs, a detailed SFH of this region would provide valuable new insights into how M82’s outflow influences star formation on larger scales. Fortunately, the photometry for the Southern Arcs is significantly cleaner and deeper than that of the ANGST fields, allowing us to determine their SFH with greater confidence.

\section{Star Formation Histories of the Southern Arcs}\label{sec:sfh}

We deduced the star formation histories (SFHs) of the Southern Arcs using the color-magnitude diagram (CMD) fitting code \texttt{MATCH} \citep{Dolphin2002, Dolphin2012, Dolphin2013}. \texttt{MATCH} determines the combination of single-burst stellar populations that best reproduces the observed CMD after accounting for photometric biases and completeness from ASTs, and fitting for dust extinction. The best-fit SFH is then established using a Poisson maximum likelihood technique \citep[see][for more details]{Dolphin2002, Dolphin2012, Dolphin2013}, which compares the number of observed stars and stars generated by the model in each CMD bin.

The CMD of the Southern Arcs does not have sufficient depth to resolve the oldest main-sequence turnoff and is therefore insensitive to the precise ages of the old population \citep[e.g.,][]{Williams2017, Weisz2014}. Additionally, since we are only interested in the younger populations, we excluded the CMD region occupied by the old RGB stars \citep[see, e.g.,][]{Lewis2015,Lazzarini2022} and focused on fitting the remaining CMD to trace the SFH of the younger population within the last $\sim500$\,Myr. The blue and red core helium burning stars (BHeB and RHeB, respectively) serve as particularly robust chronometers for recent star formation episodes, as their distinct magnitudes and colors allow precise age determinations \citep{Dohm-Palmer2001, Weisz2008, McQuinn2012}. To maximize the utility of these stars, we applied a conservative exclusion zone for the RGB (using the \texttt{exclude\_gates} functionality of \texttt{MATCH}), where we relaxed the blue cut on the RGB above the tip of the red giant branch to allow the red supergiants and the youngest RHeBs in.

The inputs for our SFH fits were: distance modulus, our choice of IMF, extinction, metallicity, binary fraction, magnitude limits, and age bins. We assumed that the younger population was at approximately the same distance as the old population and adopted a distance modulus of 27.8 \citep[TRGB distance for this field by][we also test for the effect of $\pm$0.2\,mag variations around that adopted distance modulus]{S&S2024}, set the magnitude limits to 50\% completeness limits from ASTs, used a binary fraction of 0.35, and adopted Kroupa's IMF \citep{Kroupa2001}. We used logarithmic age bins from 6.6 to 8.7 (approximately 4–500\,Myr) with a spacing of 0.15 dex between 6.6 and 7.5 and a spacing of 0.1 dex between 7.5 and 8.7. The coarser time bins at younger ages reduce the uncertainty in star formation rates (SFRs) arising from detecting fewer stars in CMD regions that correspond to faster stages of massive star evolution. This, together with the CMD exclusion of the RGB stars, ensures that we are able to isolate the distinct evolutionary histories of the younger MS and core HeB stars formed by the outflow from the older stellar-halo RGB stars. To avoid unphysical metallicities, we imposed a monotonically increasing age-metallicity relationship over our age range using the \texttt{zinc} option to capture enrichment from star formation --- a recommended strategy \citep[e.g.,][]{Weisz2008}.
To account for systematic differences in stellar evolution models used to calculate SFHs, we ran MATCH using Padova \citep{Marigo2008} isochrones with updated asymptotic giant branch (AGB) tracks \citep{Girardi2010}, MIST isochrones \citep{Dotter2016-MIST}, and updated BASTI isochrones \citep{Hidalgo2018-BASTI}. The Padova models have an [M/H] range of [$-$2.3, 0.1], the MIST models have an allowed [M/H] range of [$-$2.0, 0.5], and the BASTI models have an allowed [M/H] range of [$-$3.2, 0.4], each with a resolution of 0.1 dex. We limited the metallicities of both the oldest ($500$\,Myr) and youngest ($4$\,Myr) time bins to [$-$1.0, 0.1] and specified that the mean metallicity increases with time.

To account for extinction arising from foreground dust and dust around M82, we included foreground extinction and differential extinction in our CMD fitting process. The foreground extinction is described by $A_V$, which reddens all stars uniformly. The differential extinction is described by $dA_V$, which captures the spatially non-uniform dust distribution in M82. This extinction model is sufficient for young stars (age $<1$ Gyr) as they tend to experience a roughly top-hat differential reddening distribution \citep[eg.][]{Dolphin2003, Weisz2014, Lazzarini2022}. In \texttt{MATCH}, extinction is applied to all stars in a uniform distribution between $A_V$ and $A_V + dA_V$. Thus, $A_V$ shifts the entire CMD fainter and red-ward, whereas $dA_V$ broadens and dims the main sequence. 

We determined the best values for $A_V$ and $dA_V$ by running MATCH multiple times on a grid of $A_V$ and $dA_V$. Initially, we searched on a coarse grid that allowed $A_V$ to range between 0.1 and 1.0 in steps of 0.1 and allowed $dA_V$ to range between 0.2 and 1.2 in steps of 0.2. We then carried out a finer search in steps of 0.05 around the best fit $A_V$ and in steps of 0.1 around the best fit $dA_V$ to refine the extinction parameters. We repeated this exercise independently for all three stellar evolution models and found the best-fit combination of $A_V$ and $dA_V$ to be $0.15$ and $0.8$, respectively (see Appendix B for more details).

\subsection{Uncertainties in Star Formation Histories}

Uncertainties in our SFH calculations arise from a combination of random and systematic uncertainties, including dust. We calculated random uncertainties using a hybrid Monte Carlo (MC) process \citep{Duane1987} implemented through \texttt{MATCH}. These uncertainties scale inversely with the number of stars with more sparsely populated regions of the CMD having higher random uncertainties. The hybrid MC routine utilizes an MC algorithm that explores high-dimensional parameter spaces more efficiently than a traditional Metropolis-Hastings MC algorithm. With this, we generated 10,000 possible SFHs for the CMD from a given analysis region, where the density of the generated SFHs in parameter space is proportional to its probability density in this space. We then calculated the errors by identifying the boundaries of the region containing 68\% of the samples. This method accounts for cases where the most likely SFH has zero star formation in individual time bins, allowing the uncertainties on the star formation rate in such bins to be estimated.

To estimate systematic uncertainties due to our choice of stellar models, we recalculated the SFHs using three different sets of models: the Padova models with complete AGB tracks \citep{Marigo2008, Girardi2010}, the MIST stellar models \citep{Dotter2016-MIST}, and the BASTI models \citep{Hidalgo2018-BASTI}. Differences between these models are most pronounced in the post-main sequence phases of stellar evolution and for massive stars, primarily due to variations in modeling convection, mass loss, and rotation \citep{Conroy2013}.

We also accounted for potential systematic uncertainty from our choice of distance modulus. Literature values for the tip of the red giant branch (TRGB) distance moduli of M82 range from 27.53 to 27.95 \citep[e.g.,][]{Sakai1999, Dalcanton2009}. We ran MATCH on the Southern Arcs using distance moduli of 27.6, 27.8, and 28.0 to test how the distance modulus affects the SFH. We found that the times at which 50\% ($t_{50}$) and 90\% ($t_{90}$) of the stars in the field had formed changed by less than the size of our age bins, indicating minimal sensitivity of our results to uncertainties in distance modulus.

\begin{figure*}[htb]
    \centering
    \includegraphics[width=\linewidth]{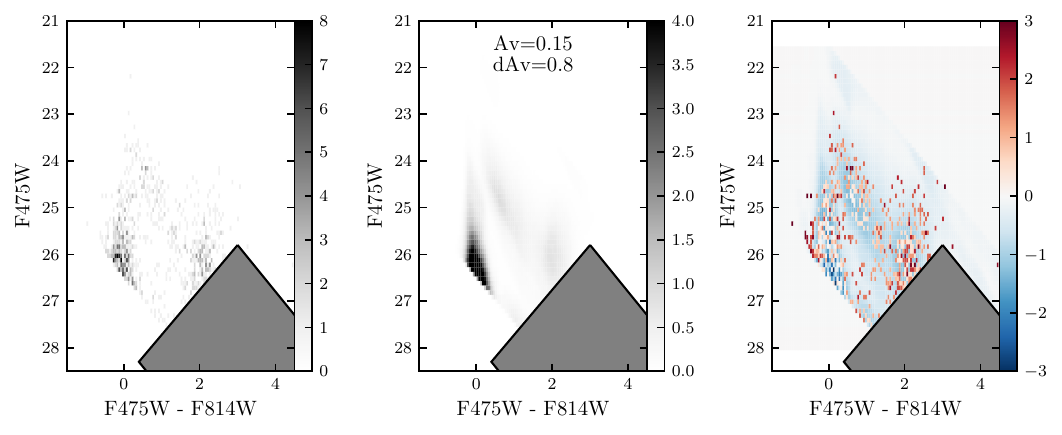}
    \caption{Southern Arcs: (\textbf{Left}) Observed Hess diagram used by \texttt{MATCH} that balances number statistics on the CMD with time resolution in the SFH; (\textbf{Center}) best-fit model Hess diagram using PADUA stellar evolution models with complete AGB tracks \citep{Marigo2008, Girardi2010}; (\textbf{Right}) residual significance Hess diagram. The color bars in the left and center panels represent the number of stars per CMD bin. In the right panel, the scaling reflects the significance of each pixel in the residual relative to the standard deviation of a Poisson distribution. The RGB mask used is shown in gray in all panels.}
    \label{fig:wfc31-cmd}
\end{figure*}

\begin{figure*}[htb]
    \centering
    \includegraphics[width=\linewidth]{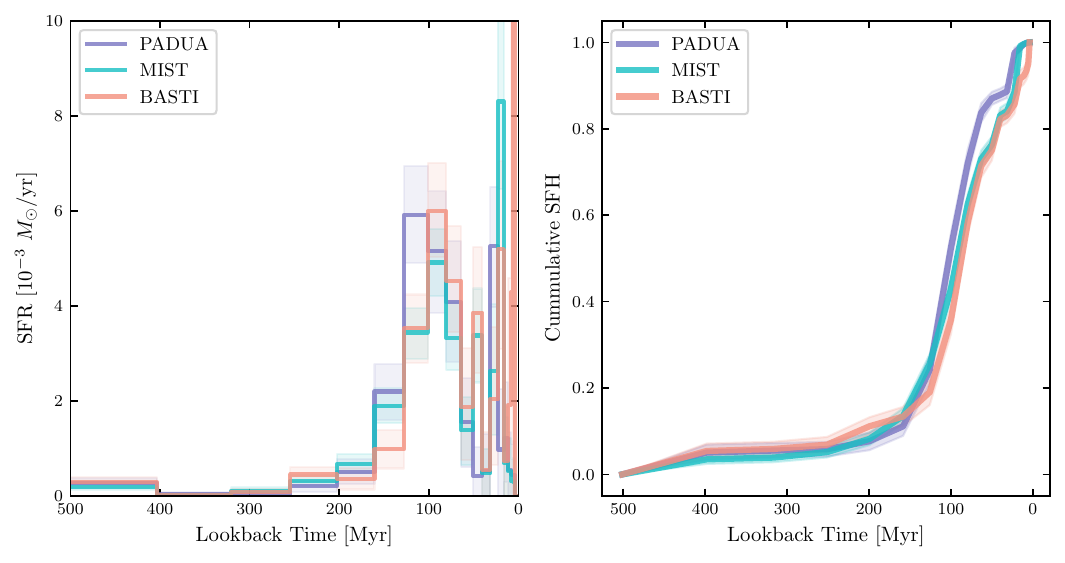}
    \caption{Best fit SFHs for the Southern Arcs using PADUA, MIST, and BASTI isochrones. Error envelopes represent the 68\% confidence interval of random uncertainties. (\textbf{Left}) Absolute star formation rate (SFR) as a function of lookback time. (\textbf{Right}) Cumulative fraction of stars formed as a function of lookback time. Star formation occurs in two epochs beginning at $\sim150$\,Myr and $\sim30$\,Myr, marked by a steepening of slopes in the cumulative SFH.}
    \label{fig:wfc31-sfh}
\end{figure*}

\subsection{Star Formation History Results}\label{sec:results}
In this section, we present the model Hess diagram (2D histogram of the CMD) and star formation histories for the Southern Arcs.

We show the quality of our Hess diagram fit for the Southern Arcs using PADUA models in Figure \ref{fig:wfc31-cmd}. The scaling of the residual Hess diagram indicates the pixel-wise deviation of the observed Hess diagram from the model relative to the standard deviation of a Poisson distribution. Overall, we find that the model reproduces the observed Hess diagram reasonably well with 95\% of pixels having a residual of $<1$ star. The best-fit model Hess diagram contains prominent main-sequence, main-sequence turn-off, and horizontal branch features that are well matched to observations in terms of density of stars, luminosities, colors, and scatter. However, we also see an under-fitting of the stars towards the top left of the RGB exclusion zone. This is likely due to old RGB stars leaking into the CMD by our conservative RGB mask that lets in the reddest He burners at the cost of a few blue RGB stars. 

The SFR of the Southern Arcs as a function of lookback time and cumulative SFH, i.e fraction of total stellar mass formed prior to a given epoch, are shown in Figure \ref{fig:wfc31-sfh}. We will mainly interpret the cumulative SFHs as they minimize many of the issues that affect interpreting absolute SFHs such as interpreting covariant SFRs in adjacent time bins and defining appropriate time resolutions \citep[e.g.,][]{H&Z2001, Dolphin2002, McQuinn2010b}. We find that $\sim85\%$ of the stellar mass in this field formed between lookback times of $\sim150$\,Myrs to $\sim70$\,Myrs after which there was a brief lull in star formation, which can be seen as a flattening in the slopes of the cumulative SFH curves. The star formation picks up again around $\sim30$\,Myrs, forming $\sim15\%$ of the stellar mass in the field. The total stellar mass formed in the last $\sim500\,$Myr is $5.3\pm0.3 \times10^5\,M_\odot$.

The SFH at ages younger than $\sim10$\,Myr is model-dependent. The PADUA and MIST models show a declining SFR toward the present day, whereas the BASTI models indicate a higher SFR at these young ages. However, the BASTI models have a limited mass range ($0.1$–$15\,M_\odot$), making SFR estimates at $\lesssim20$\,Myr uncertain \citep{Hidalgo2018-BASTI}. Additionally, significant contamination from Milky Way (MW) foreground stars in regions of the CMD occupied by post-MS stars at these ages (see Appendix A) introduces further uncertainty in the SFRs at $\lesssim10$\,Myr.

\section{Are the Bursts Spatially Coincident?}\label{sec:spat-dist}
\begin{figure*}[htb]
    \centering
    \includegraphics[width=0.95\linewidth]{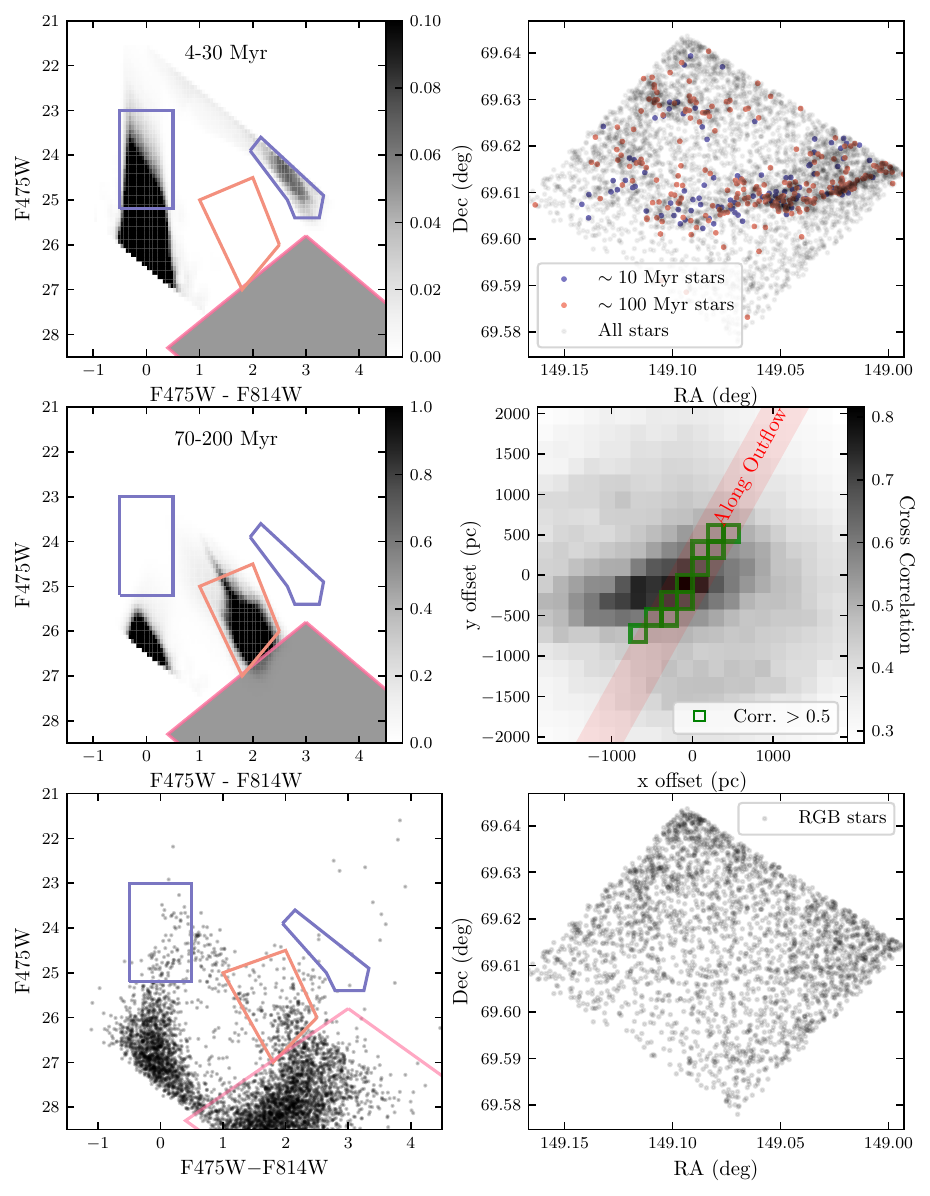}
    \caption{(\textbf{Left}) CMD selection of $\sim10$\,Myr and $\sim100$\,Myr stars from model Hess diagrams generated from PADUA models. (\textbf{Top Right}) Spatial distribution of $\sim10$\,Myr stars (blue) and $\sim100$\,Myr stars (orange). (\textbf{Middle Right}) 2D spatial cross-correlation between the $\sim10\,$Myr and $\sim100\,$Myr populations. Pixels that correspond to shifts along the direction of the outflow and having a correlation $>0.5$ are marked in green. (\textbf{Bottom Right}) RGB stars from the ancient halo population appear randomly distributed.}
    \label{fig:spat-dist}
\end{figure*}

The stars formed during the two star formation episodes, which began $\sim150$\,Myr ago and $\sim30$\,Myr ago, are spatially co-located (Fig. \ref{fig:spat-dist}). To distinguish between these populations in CMD space, we used model Hess diagrams generated by \texttt{MATCH}, with star formation rates spanning $4-30$\,Myr and $70-200$\,Myr. By identifying regions of the CMD uniquely occupied by stars from each episode, we mapped their spatial distributions (top right panel, Fig. \ref{fig:spat-dist}). These distributions reveal that both populations are concentrated along the same filamentary structures.

To quantify this spatial alignment, we computed the 2D cross-correlation of the 2D histograms representing the two stellar populations (center right panel, Fig. \ref{fig:spat-dist}), using bin sizes of $\sim200$\,pc. The central value of the cross-correlation map reflects the degree of spatial correlation at scales of $\lesssim200$ pc, where a value of 1 indicates perfect correlation, 0 indicates no correlation, and -1 indicates perfect anti-correlation. The correlation value at an offset of $(x,y)$ pc represents the degree of correlation if one histogram were shifted by $(x,y)$ pc relative to the other. We find a peak correlation of 0.81 at zero offset, with additional correlation for shifts along the filamentary structures. To estimate an upper bound on the relative sky-plane motion of these populations over a $\sim100$\,Myr timescale, we determine the maximum offset along the outflow direction where the correlation remains $>0.5$. This analysis yields a relative speed of $\sim6\,\rm{km/s}$.

\begin{figure*}[htb]
    \centering
    \includegraphics[width=\linewidth]{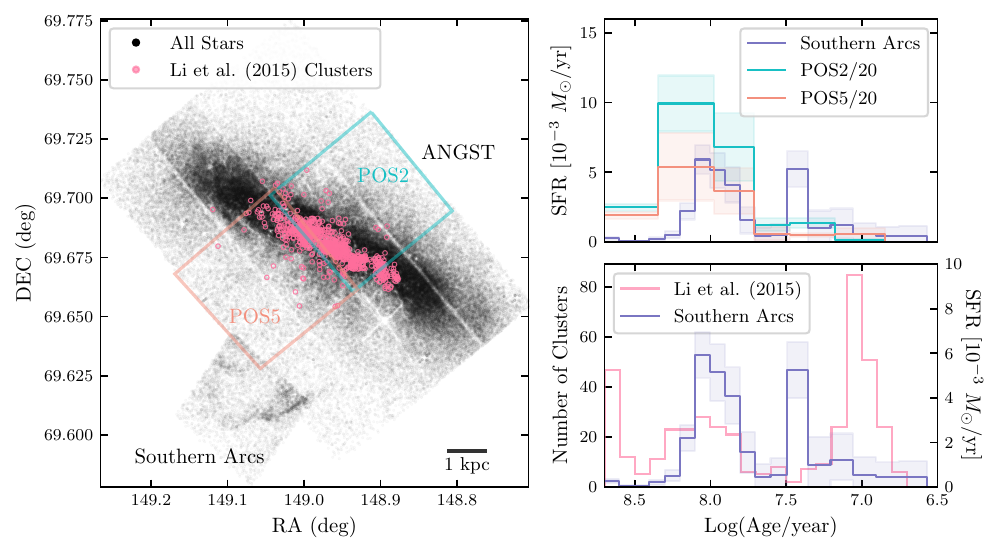}
    \caption{The SFH of the Southern Arcs appears correlated with the SFH of the central M82 disk. (\textbf{Left}) Resolved stars in the M82 field from ANGST and the Southern Arcs. Regions for which SFHs were calculated by \citet{Williams2011} (POS2 and POS5) are shown in the same plot. Locations of young massive star clusters (YMCs) studied by \citet{Li2015} are also shown for reference. (\textbf{Top Right}) SFHs of POS2 (cyan) and POS5 (orange) scaled down by a factor of 20 compared to the SFH of the Southern Arcs obtained from PADUA models (blue). POS2 and POS5 are dominated by stars from the central disk region of M82. The Southern Arcs and the central M82 disk both experienced enhanced star formation $\sim100$\,Myr ago. (\textbf{Bottom Right}) Histogram of star cluster ages from \citet{Li2015} (pink) compared to the SFH of the Southern Arcs obtained from PADUA models (blue). The M82 disk experienced multiple epochs of star cluster formation similar to the Southern Arcs.}
    \label{fig:age-compare}
\end{figure*}

\begin{figure*}[htb]
    \centering
    \includegraphics[width=\linewidth]{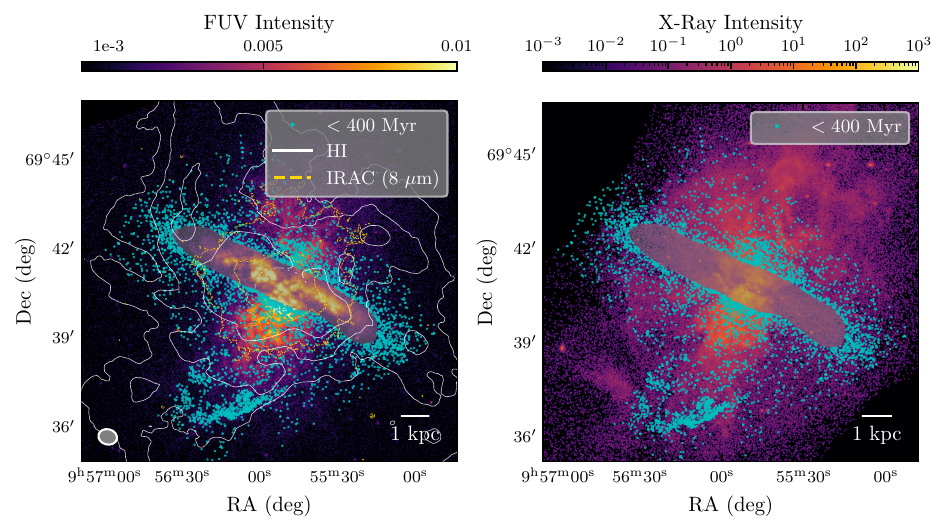}
    \caption{Multi-wavelength overview of M82's outflow and its resolved stellar populations (\textbf{Left}) The colormap represents the far-UV (FUV) intensity from GALEX (GII-36; PI: John Huchra). Most of the observed FUV photons originate from the continuum emission of the starburst regions, scattered into our line of sight by dust in the outflow. Faint FUV emission is also detected along the young stars in the Southern Arcs. HI contours \citep{deBlok2018}, tracing neutral hydrogen at levels of 0.3, 0.6, and 1.2\,Jy/beam·km/s, are shown as solid white lines. Spitzer-IRAC4 contours, which trace hot dust emission at $8\mu$m (SINGS1\_159; PI: Robert Kennicut) at levels of 1, 4, and 16 a.u (arbitrary units), are shown as dashed yellow lines. (\textbf{Right}) Archival $0.35$–$1.1\,$keV (soft) X-ray image of M82 from the Chandra X-ray Observatory.
    Resolved young stars ($\lesssim400$ Myr) from ANGST and the Southern Arcs are displayed in both panels as cyan points, with the disk of M82 masked out. In M82, soft X-ray emission originates both from the mass-loaded starburst wind itself ($T \gtrsim 10^6\,$K) and from intermediate-temperature gas ($T \sim 10^6$–$10^7\,$K) at the shocked interfaces between the hot outflow ($T \sim 10^8\,$K) and the cooler halo medium.}
    \label{fig:m82-multi-wav}
\end{figure*}

\section{Discussion}\label{sec:discussion}
In Section \ref{sec:results}, we found that the Southern Arcs formed approximately 85\% of its stellar mass between $\sim150$\,Myr and $\sim70$\,Myr ago, followed by a brief pause in star formation. The remaining 15\% of its stellar mass formed after $\sim30$\,Myr ago. \citet{S&S2024} similarly noted that older populations in the Southern Arcs align well with a $\sim100$\,Myr isochrone. The younger population we identify in this work is in addition to this $\sim100$\,Myr population. We now interpret this SFH in the context of the broader evolutionary history of M82’s stellar populations and outflow.

\subsection{Star Formation History of the M82 Disk}
The Southern Arcs’s peaks in star formation appear correlated with periods of heightened star formation in the M82 disk (Fig. \ref{fig:age-compare}). Specifically, the burst of star formation $\sim100$\,Myr ago in the Southern Arcs aligns with enhanced star formation $\sim100$\,Myr ago in POS2 and POS5, as observed by \citet{Williams2011} (top right panel of Fig. \ref{fig:age-compare}). The SFHs for POS2 and POS5, derived from the ANGST dataset by \citet{Williams2011} using a methodology similar to ours, are dominated by disk stars. Thus, the SFHs of these fields reflect the formation history of the relatively unattenuated portions of M82's disk.

The age distribution of young massive star clusters (YMCs) in the M82 disk \citep{Li2015} also shows a peak at $\sim100$\,Myr, in addition to a younger peak at $\sim10$\,Myr (bottom right panel of Fig. \ref{fig:age-compare}). These findings are consistent with \citet{Lim2013}, who identified two populations of nuclear star clusters centered around $\sim5$\,Myr and $\sim300$\,Myr. Similarly, \citet{Smith2007} found a peak in cluster ages at $\sim150$\,Myr, with ongoing cluster formation until $\sim12$–$20$\,Myr in a “fossil” starburst region located 0.5–1\,kpc northeast of the galaxy center (M82 Region B). 

It is also important to consider the findings of \citet{Schreiber2003}, who identified global starburst episodes $\sim10$\,Myr and $\sim5$\,Myr ago in M82 through near-infrared integral field spectroscopy and mid-infrared spectroscopy. These starburst events, which operated on timescales of $\sim10^6$ years, are not evident in all the SFH models of the Southern Arcs. However, as discussed in Section \ref{sec:results}, our ability to resolve star formation for lookback times $\lesssim10$\,Myr is limited by contamination from MW foreground stars in regions of the CMD that uniquely constrain such young populations. Additionally, short cadence bursts at these times can be hard to accurately resolve when fully accounting for systematics. This limitation means that a younger burst of star formation may exist in the Southern Arcs but remains undetected. Since stars form from molecular clouds, it is also plausible that the Southern Arcs had exhausted their molecular gas reservoirs by this time, thereby quenching any subsequent star formation.

In summary, the SFHs derived from the ANGST dataset and the star cluster age distributions show that the SFH of the M82 disk is correlated with the SFH of the Southern Arcs, especially when considering the $\sim100$\,Myr burst of star formation. This suggests a common mechanism linking the star formation in the Southern Arcs to the star formation in the M82 disk.

\subsection{Multi-wavelength Map of M82}
A multi-wavelength map of the M82 field, including the resolved young stars (Fig. \ref{fig:m82-multi-wav}), suggests that the starburst-driven outflow may serve as the common mechanism linking star formation across regions. 

In M82, soft X-ray emission originates from the mass-loaded starburst wind itself ($T \gtrsim 10^6\,$K), from ``intermediate" temperature gas ($T \sim 10^6$–$10^7\,$K) at the shock interface, where the hot wind ($T \sim 10^8\,$K) collides with cold ambient material in the halo, and from a variety of interactions between hot and cool phases that form arches, bow-shocks, and filaments \citep[e.g.,][]{Hoopes2003, Cooper2008, Lopez2020}. The diffuse FUV emission, which broadly traces the extent of the bi-conical outflow, primarily arises from FUV photons within the disk scattering off dust in the outflow \citep{Hoopes2005}. For additional context, we include Spitzer-IRAC4 contours, which trace hot dust emission, and HI contours \citep{deBlok2018}, which reveal the distribution of neutral hydrogen. 

The line-of-sight column densities of HI at the location of the Southern Arcs range from $3-6\times10^{20}\,\mathrm{cm}^{-2}$. \citet{deBlok2018} note that these column densities are comparable to a minimum column density threshold for star formation, which ranges from $3-10\times 10^{20}\,\mathrm{cm}^{-2}$ \citep{Schaye2004}. However, it is not clear, how this by itself can explain the star formation in the Southern Arcs given the existence of regions, for example, northeast and northwest of the M82 disk, which are rich in HI ($N_H > 6\times10^{20}\,\mathrm{cm}^{-2}$) but do not show evidence of active star formation. In other words, there is no clear overdensity that sets the Southern Arcs region apart from other regions which are similarly, if not more dense in HI. This fact together with the physical proximity of the outflow suggests that the outflow may have played a direct role in triggering star formation in the Southern Arcs.

Given these pieces of observational evidence, we explore two possible mechanisms for the formation of the Southern Arcs.

\subsection{Shock-induced Star Formation}
The starburst activity in M82 is widely attributed to its tidal interaction with M81 within the last $\sim1\,$Gyr. This interaction likely generated strong large-scale torques and increased cloud-cloud collision rates, driving gas into the galaxy’s inner regions—a process consistent with numerical simulations \citep[e.g.,][]{Noguchi1988, M&H1996}. This influx of material could have fueled intense starburst episodes, leading to the rapid formation of stars and star clusters while also launching powerful kiloparsec-scale outflows. We hypothesize that star formation in the Southern Arcs could be triggered by shocks produced by the outflows. These shocks, generally linked to some form of interaction between the hot and cool phases of the outflow or the ambient medium, could have encountered colder, denser CGM clouds and compressed the gas, leading to gravitational collapse and subsequent star formation along shock fronts.

A shock requires gas to propagate, so we expect diffuse X-ray emission to be present near the Southern Arcs \citep[e.g.,][]{Crockett2012}. Archival Chandra X-ray Observatory images spanning 0.35–1.1\,keV (soft X-rays) reveal a bi-conical outflow surrounded by a complex of arcs embedded within an almost spherical bubble around the M82 disk. Some of these diffuse X-ray emitting arcs are within a projected distance of $\sim500$\,pc from the stellar arcs (Right panel Fig. \ref{fig:m82-multi-wav}). Comparing this X-ray map with three dimensional simulations of starburst-driven galactic winds by \citet{Cooper2008}, the nearly spherical soft X-ray bubble may be the shock front formed by the fragmented shell of ambient halo gas swept up as the outflow expands into the CGM. In the simulations, such bright, bi-conical regions of soft X-ray emission interior to the swept up shell are produced by mass-loading of the hot wind by cool gas clouds driven out of the galactic disks. The complex substructure of arcs and filaments could be a result of bow shocks or oblique shocks. Bow shocks are produced by the interface of the hot wind and cool gas that has either precipitated out of the outflow or been launched from the galactic disk by the ram pressure of the hot wind.
Oblique shocks are produced by colliding bow-shocks or interactions between neighboring components of the wind \citep[e.g.,][]{Tenorio-Tagle2007}.

In this context, the $\sim100$\,Myr burst of star formation in the Southern Arcs is then easy to understand as we also see a corresponding starburst peak in the disk at $\sim100$\,Myr. Shocks from the outflow resulting from the starburst event $\sim100$\,Myr ago could have overrun dense CGM clouds, triggering star formation in the clouds.  As discussed in Section \ref{sec:intro}, this mode of star formation has been reproduced in simulations \citep{Gardner2017, Kinoshita2021a} and is analogous to AGN-jet-induced star formation observed in the inner optical filament of Cen A’s CGM, where diffuse X-ray emission from shock-heated gas is also detected in the vicinity \citep{E&K2004}. The mechanism is also analogous to that suggested for the formation of ``The Cap," a feature located approximately 11\,kpc north of the M82 disk plane and thought to have formed from an interaction between M82's northern outflow and surrounding clouds \citep[e.g.,][]{D&B1999,Lehnert1999, Strickland2004, Hoopes2005}. Although The Cap is a weak UV source \citep{Hoopes2005} and is associated with diffuse X-ray emission \citep{Strickland2004}, it remains unclear whether it contains any stellar population.

\begin{figure*}
    \centering
    \includegraphics[width=\linewidth]{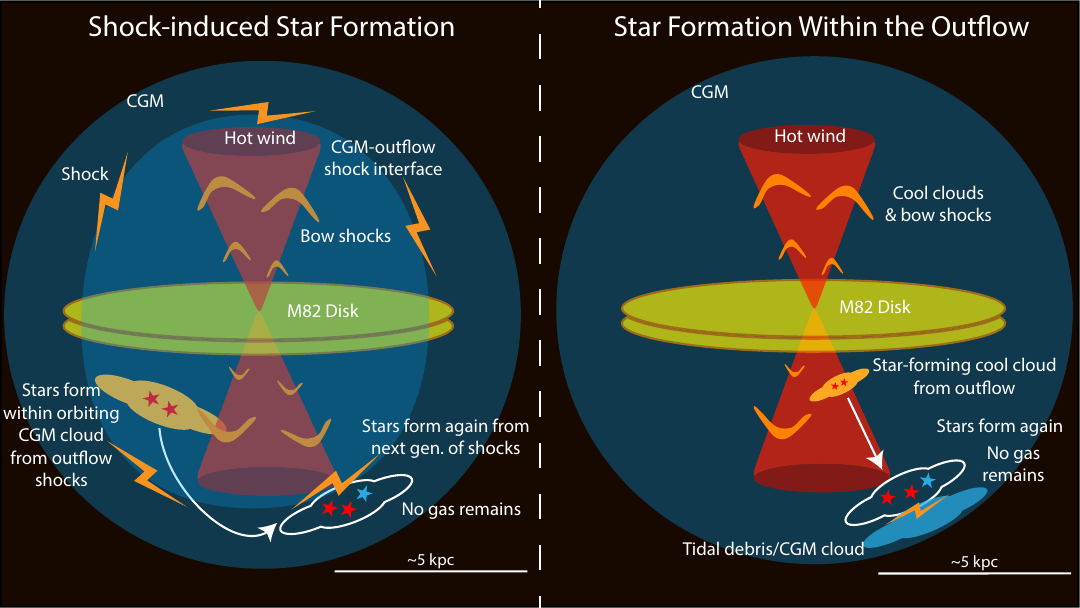}
    \caption{Schematic illustration of the two mechanisms proposed for the formation of the Southern Arcs. (\textbf{Left}) Multiple generations of shocks from the outflow compress a cool, dense CGM cloud, triggering multiple epochs of star formation. (\textbf{Right}) Stars form within cool clouds associated with the outflow. The stars and clouds co-move ballistically. An encounter with a second CGM cloud or tidal debris triggers a second burst of star formation in the original clouds.}
    \label{fig:illustration}
\end{figure*}

The subsequent co-evolution of the disk and the Southern Arcs after the $\sim100$\,Myr burst, however, is not straightforward. The second episode of star formation in the Southern Arcs, beginning at $\sim30$\,Myr, does not coincide with the second peak in star cluster ages at $\sim10$\,Myr reported by \citet{Li2015} (bottom right panel of Fig. \ref{fig:age-compare}). However, star cluster age distributions across different regions of the M82 disk reveal varying levels of starburst activity at different times over the past 100\,Myr. Moreover, the optical catalog of star clusters used by \citet{Li2015} is at least $87\%$ incomplete due to the high levels of dust extinction within the disk (see Section \ref{subsec:future}). It is therefore plausible that a weaker, shorter-duration starburst occurred around $\sim30\,$Myr ago—one not captured in the cluster age distribution, but nonetheless capable of launching outflow shocks that compressed CGM clouds a second time. This scenario (left panel of Fig. \ref{fig:illustration}) could explain the formation of a second population of stars in the Southern Arcs, nearly co-spatial with the first, as discussed in Section \ref{sec:spat-dist}. A notable implication of this picture is that both the $\sim100\,$Myr and $\sim30\,$Myr stellar populations would share similar metallicities, having formed from the same reservoir of CGM material.

In the scenario proposed above, bursty star formation episodes in the disk result in bursty outflows, which in turn trigger multiple epochs of star formation in the progenitor clouds of the Southern Arcs through shocks. If this were indeed the case, the $\sim70$–$80\,$Myr separation between peaks in SFR measured in this work would represent the first observational measurement of the cadence of a starburst outflow. However, there exists some physical challenges with the above scenario, which we highlight below.

In Section \ref{sec:spat-dist}, we found that stars formed during the $\sim150\,$Myr and $\sim30\,$Myr bursts are co-spatial on scales of at least $200\,$pc. Additionally, we calculated the relative on-sky motion between these two populations to be $\lesssim6\,$km/s, indicating minimal relative motion over the last $\sim100\,$Myr— a striking result. If we assume that the velocity dispersion of the progenitor clouds of the Southern Arcs is $\lesssim6\,$km/s, we find that their thermal temperature is $\lesssim3000\,$K (assuming a neutral, solar composition). This is not a particularly unexpected value as cold gas clouds with $T\lesssim10^4\,$K have been observed in M82's CGM, traced through HI and molecules such as CO \citep[e.g.,][]{Walter2002, Leroy2015, Martini2018}. However, given that winds and radiation from the $\sim100\,$Myr population could disperse or displace the remaining gas, it is unclear how the clouds could maintain coherence over a $\sim70\,$Myr timescale and sustain fuel for a subsequent, spatially overlapping burst.

A second challenge lies in the dynamics of the progenitor clouds. At the location of the Southern Arcs, HI gas is moving radially toward us at $\sim100\,$km/s \citep[Moment-1 maps;][]{Martini2018}. If the Southern Arcs are associated with this HI and share its line-of-sight velocity, they would take approximately $\sim300\,$Myr to complete one full circular orbit. The observation that outflow-induced star formation has potentially occurred in the same molecular clouds over a span of $\sim100\,$Myr suggests that the clouds have remained within the influence of outflow-driven shocks for about a third of its orbit. This persistence raises an interesting question: how can the outflow continue to affect the same region for such an extended period?

A possible resolution lies in the nearly spherical geometry of the soft X-ray emission (Right panel Fig. \ref{fig:m82-multi-wav}). As discussed earlier, this spherical bubble may be the shock front formed by the shell of ambient halo gas swept up as the outflow expands into the CGM- similar to what is seen in simulations by \citet{Cooper2008}. This raises the possibility that the spatial influence of outflow-driven shocks may extend more broadly than the apparent bi-conical morphology of hot gas and dust, potentially explaining the long-duration interaction with the Southern Arcs' molecular clouds. 

However, it is important to note that the simulations by \citet{Cooper2008} were limited in spatial extent, reaching only $\sim500\,$pc from the mid-plane—much smaller than the $\sim5\,$kpc distance to the Southern Arcs. It is therefore plausible that the shock front becomes weaker or more diffuse at larger distances. Outside the bi-cone, shocks may be less coherent and less capable of sustaining strong compression. These considerations may be related to the nature of the $\sim30\,$Myr starburst episode in the Southern Arcs, which was less intense than the $\sim100\,$Myr burst—potentially reflecting triggering under weaker or more marginal conditions outside the bi-cone.

\subsection{Star Formation Within the Outflow}
We now propose a second possible mechanism for the formation of the Southern Arcs. Following the starburst episode $\sim100\,$Myr ago in the M82 disk and corresponding launch of the outflow, stars could have directly formed within the cooler phases of the outflow that reached densities above the star-forming threshold.

Much work has been done in understanding the origin and fate of cool gas clouds ($T\lesssim10^4\,$K) within starburst outflows. The classical picture is that these clouds are entrained from the host ISM and accelerated by the ram pressure of the hot wind \citep[e.g.,][]{Veilleux2005}. However, it is unclear how these cool ISM clouds can survive accelerations to large velocities and scales without being shredded by the hot wind and incorporated into the flow \citep[e.g.,][]{Zhang2017}. Recent work suggests a complex interplay between the hot and cool phases, where the cool clouds exchange energy and momentum with the background hot wind material \citep[e.g.,][]{Thompson2016, G&O2018, G&O2020, Fielding&Bryan2022, Gronke2022}. Specifically, radiative cooling of the warm gas formed from the mass loading of the flow by the shredded cool clouds can seed the formation of more cool clouds. These clouds, if they are large enough, can grow and survive as cooling wins out over turbulent shredding.

In M82, there exists a complex structure of H$\alpha$ emitting filaments and arches kiloparsecs from the disk \citep[e.g.,][]{Lopez2025}. These H$\alpha$ features are co-spatial with X-ray emitting features and can be explained as a result of shock interactions between the hot wind and the cool clouds \citep[e.g.,][]{S&B1998, Cooper2008, Okon2024}. Cool clouds in M82's outflow can also be traced through 3.3$\mu$m polycyclic aromatic hydrocarbon (PAH) emission. This spectral feature traces small dust grains and suggests the launching and survival of cold gas through hierarchically structured plumes \citep{Fisher2025, Villanueva2025}. It is then possible that the $\sim100\,$Myr stars in the Southern Arcs could have been born in such cool clouds associated with the outflow $\sim100\,$Myr ago. Assuming these cool clouds are traveling ballistically, the stars formed would inherit the velocity of the clouds and co-move on ballistic trajectories.

If such clouds containing populations of co-moving stars further encountered a second CGM cloud or piece of tidal debris from the interaction with M81, the resulting shock or compression could result in a second generation of weaker star formation within the same clouds. These new stars, formed $\sim30\,$Myr ago, would be roughly co-spatial with the first generation. As the $\sim100\,$Myr population formed from metal-enriched, outflowing material, the new $\sim30\,$Myr population could potentially have a different composition- having formed from an interaction between cool clouds born out of the outflow and pieces of tidal debris. The gas dynamics of such a scenario (right panel of Fig. \ref{fig:illustration}) is clearly complex and difficult to model and depends on a number of key factors. We therefore highlight some important physical considerations.

As in the shock-induced star formation picture, the above scenario suffers from a similar challenge- how do the cold clouds in the outflow maintain coherence and survive over long periods? While it is difficult to quantify all the physical properties of the Southern Arcs' progenitor clouds today, a combination of simulations and observations of other clouds in M82 may hold clues.

It is well established through simulations that if cool gas clouds exceed a critical size, they can grow rather than be destroyed: the rate at which they cool and accrete hot wind material can outpace the rate at which they are shredded by turbulent mixing \citep[e.g.,][]{Thompson2016, G&O2018, G&O2020, Li2020, Fielding&Bryan2022}. For instance, \citet{Tan&Fielding2024} find that this minimum size is $\sim5\,$pc in their fiducial models. If the present-day size of the Southern Arcs ($\gtrsim 1\,$kpc) reflects the size of their progenitor clouds, then those clouds would have been significantly larger than this critical threshold—suggesting they could have grown in mass and survived for extended periods. Similarly, \citet{Lopez2025} identify cold clouds ranging from 14–110\,pc in size, also above the theoretical survival thresholds.

While minimum cloud column densities ($N_{H,cl}$) are another metric used to predict cloud survival, we lack constraints on the original column densities of the progenitor clouds. The values measured by \citet{deBlok2018} are line-of sight values and reflect present-day column densities- there may be projection effects or the clouds may have dispersed. Nevertheless, \citet{Lopez2025} provide $N_{H,cl}$ measurements for their clouds that may benefit comparison. They find values ranging from $N_{H,cl} \approx 10^{20}-10^{21}\,\mathrm{cm}^{-2}$, well above the minimum value derived assuming nominal values.

A second physical consideration arises: if outflow-driven star formation can occur under the right conditions, why don’t we observe similar star formation triggered by younger starburst episodes? One possibility is that the $\sim100\,$Myr starburst was uniquely intense and prolonged, as supported by SFH measurements of both the disk and the Southern Arcs. This exceptional burst may have launched especially massive and dense cool clouds—clouds capable of growing large, surviving long enough, and capable of forming stars. The presence of abundant tidal debris in the direction of M81, into which the outflow was directed, may have further enhanced the probability of triggering further star formation. However, it is also plausible that more recent starbursts have produced similar effects, but, being closer to the disk, they are hidden from view due to heavy dust obscuration in the outflow. Indeed, this scenario could be connected to the widespread distribution of young and intermediate-age stars observed throughout M82’s halo in the ANGST survey.

\subsection{Observational Tests}
The two scenarios we propose for the formation of the Southern Arcs involve highly non-linear, multi-phase gas physics that is inherently difficult to model. Each depends on a range of physical parameters—such as the detailed starburst history of M82, the efficiency of supernova feedback, wind mass-loading, the galaxy’s dynamical mass distribution, and the nature of its environment, including the composition, density, and morphology of CGM clouds. Many of these factors are challenging to constrain observationally, making it difficult to uniquely favor one scenario over the other based on theoretical grounds alone.

Nevertheless, we suggest a promising observational avenue for distinguishing between these scenarios: measuring the metallicities of stars from the two stellar populations in the Southern Arcs. If the stars from the two populations have similar metallicities, a scenario where shocks compressed the same cloud multiple times may be favorable. If the metallicities are significantly different, the second scenario of stars forming directly within the outflow and involving multiple clouds may be more likely. While both scenarios have their own challenges and uncertainties, such metallicity measurements offer a concrete path forward to better constrain the physical origin of the Southern Arcs.

\subsection{Alternative Scenarios in the Literature}
Alternative scenarios for the formation of the Southern Arcs have been explored by both \citet{Davidge2008} and \citet{S&S2024}. The first hypothesis suggests that these arcs are a tidal dwarf galaxy (TDG)—a remnant of a dwarf galaxy that was accreted into M82’s halo. However, the correlation of the Southern Arcs’ SFH with the SFH of the central disk and the star cluster age distribution challenges this interpretation. It is unlikely that the SFH of an independent dwarf galaxy accreted into the halo would mirror that of M82’s disk without a shared influence, such as the outflow. While tidal instabilities occurring $\sim 100$\,Myr ago could theoretically affect both systems, this scenario seems improbable. Furthermore, there is no evidence of an underlying RGB population associated with these arcs (bottom-right panel Fig. \ref{fig:spat-dist}). This absence of RGB stars—expected in a disrupted dwarf galaxy—indicates that this hypothesis is unlikely.

The second scenario proposes that these arcs formed in situ due to tidal interactions, similar to other star-forming tidal features in the M81 Group, such as Holmberg IX, the Garland, or the Arp Loop. However, unlike these features, which exhibit HI overdensities, the Southern Arcs field is comparatively sparse in HI, as noted by both \citet{Davidge2008} and \citet{S&S2024}. This lack of HI further weakens the likelihood of this scenario. We thus conclude that neither hypothesis adequately explains the formation of the stellar arcs, reinforcing the idea that the outflow played a critical role in shaping the Southern Arcs’ star formation history.

\subsection{Future Work}\label{subsec:future}
Since the launch of the James Webb Space Telescope (JWST), there has been growing interest in M82 and the mechanisms driving its powerful starburst wind. JWST observations with the Near Infrared Camera (NIRCam) and Mid-Infrared Instrument (MIRI) have targeted the central $\sim900$\,pc of M82, enabling detailed studies of the structures and physical processes involved in launching molecular gas, traced through features such as the 3.3$\mu$m PAH emission \citep{Bolatto2024, Villanueva2025, Fisher2025}. From this same dataset, \citet{Levy2024} identified 1183 new star cluster candidates, further emphasizing the complexity and richness of M82’s central region.

These observations will soon be complemented by upcoming deep NIRCam imaging from JWST Cycle 3 program GO-5145 (PI: Smercina), which will provide resolved stellar population maps across M82's disk and minor axis, to radii of 7\,kpc and 3.5 \,kpc, respectively. While this new dataset will not cover the Southern Arcs, it will penetrate regions of high dust extinction in the central few kiloparsecs, resolving young stellar populations previously hidden in the optical. These measurements will improve our understanding of the drivers and time evolution of the starburst wind. Additionally, the ability to measure detailed star formation histories (SFHs)—including the ancient red clump (RC)—across a significant portion of the halo will allow us to assess where and how the starburst outflow has influenced star formation beyond the disk. Wide-area resolved stellar population maps from future surveys with the Nancy Grace Roman Space Telescope, covering regions not accessible with JWST, will further enhance this picture.

Together, this study and the promise of the upcoming JWST data underscore the need for more sophisticated hydrodynamical simulations of M82’s outflow. Such models should not only capture multiple starburst  cycles, but also the conditions under which outflows could trigger halo star formation- be it through shocks or interactions between multiple cool clouds.

\section{Summary}\label{sec:summary}
We used HST/WFC3 imaging to resolve stars within arc-like features located $\sim5$\,kpc south of the M82 disk. These “Southern Arcs” appear to be extensions of structures identified in the ANGST survey that are spatially associated with M82’s starburst-driven outflow. Using the CMD-fitting software \texttt{MATCH}, we derived detailed star formation histories (SFHs) over the last $\sim500$\,Myr and identified at least two distinct bursts of star formation in the Southern Arcs. Roughly 85\% of their stellar mass formed between $\sim150$ and $\sim70$\,Myr ago, followed by a brief quiescent period, after which another 15\% formed in a burst beginning around $\sim30$\,Myr ago. The bursts are also co-spatial on scales of at least $\sim200\,$pc.

The timing of $\sim100\,$Myr burst correlates with features in both the M82 disk SFH and the age distribution of star clusters, suggesting that the Southern Arcs formed stars as a direct result of M82’s starburst outflow. We thus propose two physical mechanisms through which the outflow could influence star formation in the Southern Arcs. In the first, shocks generated as hot outflowing gas interacts with cooler halo gas compress CGM clouds, triggering gravitational collapse. In this case, bursty star formation in the M82 disk drives episodic outflows, which in turn induce the multiple star formation episodes observed in the Southern Arcs. In the second scenario, stars form directly within large, cool clouds associated with the outflow and co-move with them ballistically. As these clouds travel through the halo, subsequent interactions with surrounding tidal debris may trigger additional episodes of star formation.

Through the course of this work, we considered multiple scenarios for the formation of the Southern Arcs and ultimately settled on the two most plausible—though each still presents significant physical challenges. It remains possible that neither scenario fully captures the true formation mechanism of the Arcs. However, measuring the metallicities of stars from the two stellar populations may offer a way to distinguish between them, providing valuable insight into the actual origin of the Arcs.

\begin{acknowledgments}
We thank the anonymous referee for their insightful comments, which improved this work. We also acknowledge useful conversations with Mateusz Ruszkowski. 
This work was partly supported by HST grant GO-16185 provided by NASA through a grant from the Space Telescope Science Institute, which is operated by the Association of Universities for Research in Astronomy, Inc., under NASA contract NAS5-26555. We acknowledge support from the National Science Foundation through grant NSF-AST 2007065. AS is supported by NASA through the Hubble Fellowship grant HST-HF2-51567 awarded by STScI.
AM acknowledges support from the FONDECYT Regular grant 1212046, from the ANID BASAL project FB210003, and funding from the HORIZON-MSCA-2021-SE-01 Research and Innovation Programme under the Marie Sklodowska-Curie grant agreement number 101086388. 

This research is based on observations made with the NASA/ESA \textit{Hubble Space Telescope} General Observer program 16185. This research is partly based on observations made with the \textit{Galaxy Evolution Explorer}, obtained from the MAST data archive at the Space Telescope Science Institute. This work is based in part on observations made with the \textit{Spitzer Space Telescope}, which was operated by the Jet Propulsion Laboratory, California Institute of Technology under a contract with NASA. This research uses data obtained from the Chandra Data Archive provided by the Chandra X-ray Center (CXC). This work has made use of data from the European Space Agency (ESA) mission
{\it Gaia} (\url{https://www.cosmos.esa.int/gaia}), processed by the {\it Gaia}
Data Processing and Analysis Consortium (DPAC,
\url{https://www.cosmos.esa.int/web/gaia/dpac/consortium}). Funding for the DPAC
has been provided by national institutions, in particular the institutions
participating in the {\it Gaia} Multilateral Agreement.

\end{acknowledgments}

%

\vspace{5mm}
\facilities{HST, GALEX, Spitzer (STIS), VLA (NRAO), Mikulski Archive for Space Telescopes (MAST)}


\software{numpy \citep{numpy}, matplotlib \citep{matplotlib}, BEAST \citep{Gordon2016}, astropy \citep{astropy}, DOLPHOT \citep{Dolphin2000PHOT}, MATCH \citep{Dolphin2002, Dolphin2012, Dolphin2013,Dolphin2016}}



\appendix\label{appendix:mw-foreground}

\section{Milky Way Foreground}
\begin{figure*}[htb]
    \centering
    \includegraphics[width=\linewidth]{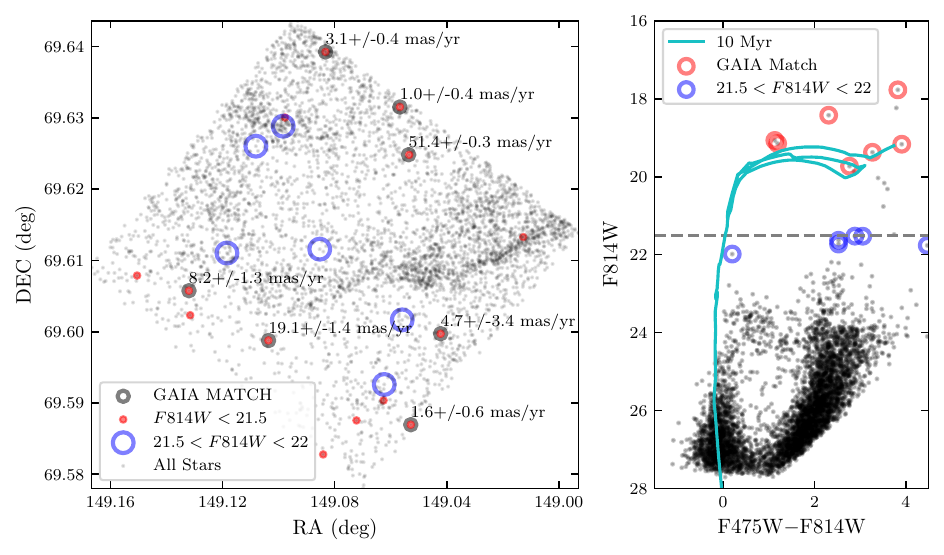}
    \caption{(\textbf{Left}) Spatial distribution of stars in the Southern Arcs field. MW foreground stars identified using Gaia DR3 \citep{GAIA-mission, GAIA-DR3} are shown with their proper motions and are circled in gray. (\textbf{Right}) The CMD of stars in the Southern Arcs field with a $10\,$Myr PARSEC isochrone over-plotted. Stars with F814$<21.5$ are likely MW foreground}
    \label{fig:m82-gaia}
\end{figure*}

Of the 14 stars with F814W$<21$ in the Southern Arcs field, we identified 7 of these to be MW foreground stars with a non-zero proper motion (Fig. \ref{fig:m82-gaia}). These stars occupy regions of the CMD that are typically populated by $\lesssim10\,$Myr post MS stars. The spatial locations of the remaining 7 stars are more or less random and broadly do not correspond to the arc-like features seen in this field. Stars just below this magnitude range ($21.5<$F814W$<22$) appear distributed near these arc-like features. We thus only consider stars with F814W$<21.5$ in this work.

\section{Dust Parameters}\label{appendix:dust}

\begin{figure*}[htb]
    \centering
    \includegraphics[width=\linewidth]{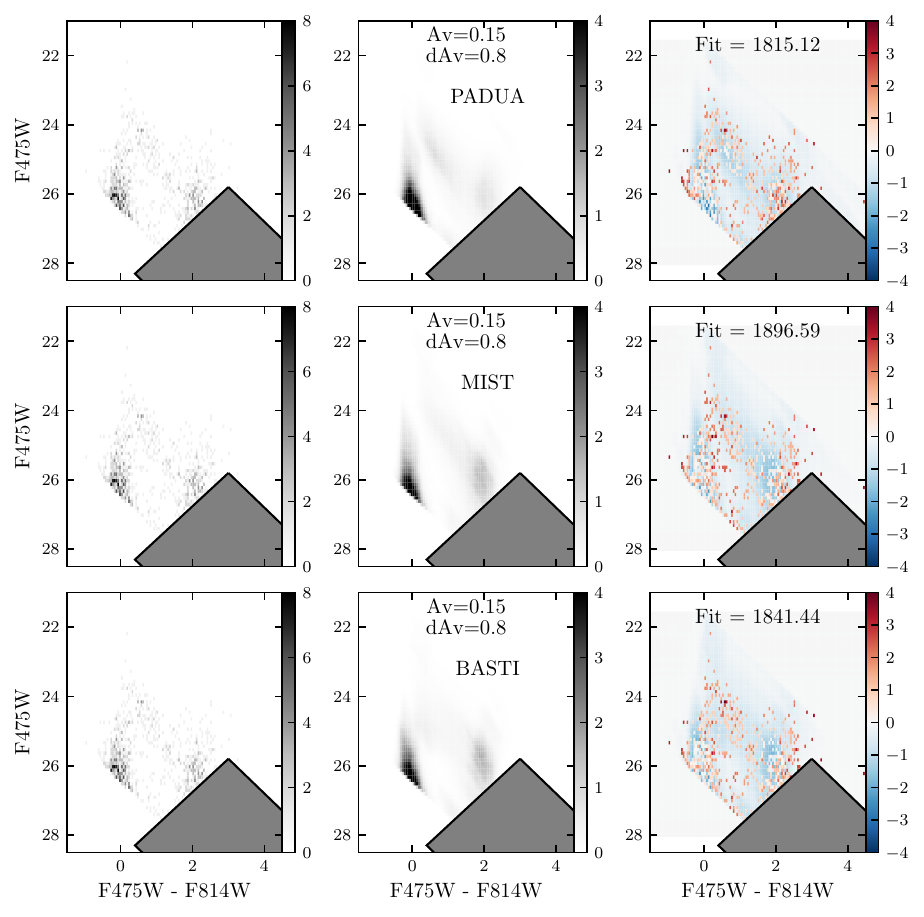}
    \caption{Southern Arcs: (\textbf{Left}) Observed Hess diagram; (\textbf{Center}) best-fit model Hess diagram using PADUA (\textbf{top row}), MIST (\textbf{middle row}) and BASTI (\textbf{bottom row}) stellar evolution models; (\textbf{Right}) Corresponding residual significance Hess diagrams}
    \label{fig:all-cmds}
\end{figure*}

In Figure \ref{fig:all-cmds}, we show the quality of our Hess diagram fits for the Southern Arcs using the PADUA, MIST, and BASTI stellar evolution models. During the dust parameter optimization process, we found that a differential extinction value of $dA_V = 1.0$ provided a marginally better statistical fit to the CMDs. However, the resulting model CMDs exhibited overly smeared features, particularly in the RHeB and MS strips, which deviated noticeably from the morphology seen in the observed CMD. To balance statistical fit quality with visual and physical realism, we adopted the next-best set of dust parameters—$(A_V, dA_V) = (0.15, 0.8)$—which produced model CMDs with more realistic stellar evolutionary features. Importantly, the choice between these dust configurations has only a minor effect on the inferred star formation histories (SFHs), with key age percentiles such as $t_{50}$ and $t_{90}$ differing by less than the width of our time bins on average.


\bibliography{main-v3}{}
\bibliographystyle{aasjournal}



\end{document}